\documentclass[12pt]{article}
\usepackage{graphicx}
\usepackage{amsmath}
\usepackage{xspace}
\usepackage{ifthen}

\newcommand\pubdate{November 14, 2014}

\newcommand{\paperAuthor}{Alberto Lusiani}
\newcommand{\paperInst}{INFN and Scuola Normale Superiore, Pisa, Italy}

\def\Title#1{\begin{center} {\Large #1 } \end{center}}
\def\Author#1{\begin{center}{ \sc #1} \end{center}}
\def\Address#1{\begin{center}{ \it #1} \end{center}}

\newcommand\pubblock{\rightline{\begin{tabular}{l} \pubnumber\\
      \pubdate  \end{tabular}}}
\renewcommand\pubblock{\rightline{\pubdate}}
\newenvironment{Abstract}{\begin{quotation}  }{\end{quotation}}
\newenvironment{Presented}{\begin{quotation} \begin{center} 
      PRESENTED AT\end{center}\bigskip 
      \begin{center}\begin{large}}{\end{large}\end{center} \end{quotation}}

\def\beq{\begin{equation}}
\def\eeq#1{\label{#1}\end{equation}}
\def\eeqn{\end{equation}}

\def\beqa{\begin{eqnarray}}
\def\eeqa#1{\label{#1}\end{eqnarray}}
\def\eeqan{\end{eqnarray}}

\let\bar=\overbar

\def\Dslash{\not{\hbox{\kern-4pt $D$}}}
\def\dslash{\not{\hbox{\kern-2pt $\del$}}}

\def\BR{\mbox{\rm BR}}

\def\msb{{\bar{\ssstyle M \kern -1pt S}}}

\input papermacros.tex
\htdef{BrStrangeVal}{%
\begin{ensuredisplaymath}
\htuse{Gamma10.gn} = \htuse{Gamma10.td}
\end{ensuredisplaymath}
 & \ensuremath{(0.6955 \pm 0.0096) \cdot 10^{-2}} \\
\begin{ensuredisplaymath}
\htuse{Gamma16.gn} = \htuse{Gamma16.td}
\end{ensuredisplaymath}
 & \ensuremath{(0.4331 \pm 0.0149) \cdot 10^{-2}} \\
\begin{ensuredisplaymath}
\htuse{Gamma23.gn} = \htuse{Gamma23.td}
\end{ensuredisplaymath}
 & \ensuremath{(0.0630 \pm 0.0220) \cdot 10^{-2}} \\
\begin{ensuredisplaymath}
\htuse{Gamma28.gn} = \htuse{Gamma28.td}
\end{ensuredisplaymath}
 & \ensuremath{(0.0419 \pm 0.0216) \cdot 10^{-2}} \\
\begin{ensuredisplaymath}
\htuse{Gamma35.gn} = \htuse{Gamma35.td}
\end{ensuredisplaymath}
 & \ensuremath{(0.8378 \pm 0.0123) \cdot 10^{-2}} \\
\begin{ensuredisplaymath}
\htuse{Gamma40.gn} = \htuse{Gamma40.td}
\end{ensuredisplaymath}
 & \ensuremath{(0.3680 \pm 0.0103) \cdot 10^{-2}} \\
\begin{ensuredisplaymath}
\htuse{Gamma44.gn} = \htuse{Gamma44.td}
\end{ensuredisplaymath}
 & \ensuremath{(0.0124 \pm 0.0204) \cdot 10^{-2}} \\
\begin{ensuredisplaymath}
\htuse{Gamma53.gn} = \htuse{Gamma53.td}
\end{ensuredisplaymath}
 & \ensuremath{(0.0222 \pm 0.0202) \cdot 10^{-2}} \\
\begin{ensuredisplaymath}
\htuse{Gamma128.gn} = \htuse{Gamma128.td}
\end{ensuredisplaymath}
 & \ensuremath{(0.0153 \pm 0.0008) \cdot 10^{-2}} \\
\begin{ensuredisplaymath}
\htuse{Gamma130.gn} = \htuse{Gamma130.td}
\end{ensuredisplaymath}
 & \ensuremath{(0.0048 \pm 0.0012) \cdot 10^{-2}} \\
\begin{ensuredisplaymath}
\htuse{Gamma132.gn} = \htuse{Gamma132.td}
\end{ensuredisplaymath}
 & \ensuremath{(0.0093 \pm 0.0015) \cdot 10^{-2}} \\
\begin{ensuredisplaymath}
\htuse{Gamma151.gn} = \htuse{Gamma151.td}
\end{ensuredisplaymath}
 & \ensuremath{(0.0410 \pm 0.0092) \cdot 10^{-2}} \\
\begin{ensuredisplaymath}
\htuse{Gamma801.gn} = \htuse{Gamma801.td}
\end{ensuredisplaymath}
 & \ensuremath{(0.0037 \pm 0.0014) \cdot 10^{-2}} \\
\begin{ensuredisplaymath}
\htuse{Gamma802.gn} = \htuse{Gamma802.td}
\end{ensuredisplaymath}
 & \ensuremath{(0.2923 \pm 0.0068) \cdot 10^{-2}} \\
\begin{ensuredisplaymath}
\htuse{Gamma803.gn} = \htuse{Gamma803.td}
\end{ensuredisplaymath}
 & \ensuremath{(0.0411 \pm 0.0143) \cdot 10^{-2}} \\
\begin{ensuredisplaymath}
\htuse{Gamma822.gn} = \htuse{Gamma822.td}
\end{ensuredisplaymath}
 & \ensuremath{(0.0001 \pm 0.0001) \cdot 10^{-2}} \\
\begin{ensuredisplaymath}
\htuse{Gamma833.gn} = \htuse{Gamma833.td}
\end{ensuredisplaymath}
 & \ensuremath{(0.0001 \pm 0.0001) \cdot 10^{-2}}}%
\htdef{BrStrangeTotVal}{%
\begin{ensuredisplaymath}
\htuse{Gamma110.gn} = \htuse{Gamma110.td}
\end{ensuredisplaymath}
 & \ensuremath{(2.8816 \pm 0.0470) \cdot 10^{-2}}}%

\htquantdef{B_tau_s_fit}{B_tau_s_fit}{}{2.882 \pm 0.047}{2.882}{0.047}%
\htquantdef{B_tau_s_unitarity}{B_tau_s_unitarity}{}{\ensuremath{(2.980 \pm 0.100) \cdot 10^{-2}}}{2.980\cdot 10^{-2}}{0.100\cdot 10^{-2}}%
\htquantdef{B_tau_VA}{B_tau_VA}{}{\ensuremath{0.6181 \pm 0.0010}}{0.6181}{0.0010}%
\htquantdef{B_tau_VA_fit}{B_tau_VA_fit}{}{61.81 \pm 0.10}{61.81}{0.10}%
\htquantdef{B_tau_VA_unitarity}{B_tau_VA_unitarity}{}{\ensuremath{0.61910 \pm 0.00079}}{0.61910}{0.00079}%
\htquantdef{Be_fit}{Be_fit}{}{\ensuremath{0.17817 \pm 0.00041}}{0.17817}{0.00041}%
\htquantdef{Be_from_Bmu}{Be_from_Bmu}{}{\ensuremath{0.17882 \pm 0.00041}}{0.17882}{0.00041}%
\htquantdef{Be_from_taulife}{Be_from_taulife}{}{\ensuremath{0.17778 \pm 0.00033}}{0.17778}{0.00033}%
\htquantdef{Be_unitarity}{Be_unitarity}{}{\ensuremath{0.17916 \pm 0.00097}}{0.17916}{0.00097}%
\htquantdef{Be_univ}{Be_univ}{}{17.815 \pm 0.023}{17.815}{0.023}%
\htquantdef{Bmu_by_Be_th}{Bmu_by_Be_th}{}{\ensuremath{0.9725594 \pm 0.0000048}}{0.9725594}{0.0000048}%
\htquantdef{Bmu_fit}{Bmu_fit}{}{\ensuremath{0.17392 \pm 0.00040}}{0.17392}{0.00040}%
\htquantdef{Bmu_from_taulife}{Bmu_from_taulife}{}{\ensuremath{0.17290 \pm 0.00032}}{0.17290}{0.00032}%
\htquantdef{Bmu_unitarity}{Bmu_unitarity}{}{\ensuremath{0.17490 \pm 0.00097}}{0.17490}{0.00097}%
\htquantdef{BR_eta_2gam}{BR_eta_2gam}{}{0.3941}{0.3941}{0}%
\htquantdef{BR_eta_3piz}{BR_eta_3piz}{}{0.3268}{0.3268}{0}%
\htquantdef{BR_eta_charged}{BR_eta_charged}{}{0.2810}{0.2810}{0}%
\htquantdef{BR_eta_neutral}{BR_eta_neutral}{}{0.7212}{0.7212}{0}%
\htquantdef{BR_eta_pimpippiz}{BR_eta_pimpippiz}{}{0.2292}{0.2292}{0}%
\htquantdef{BR_f1_2pip2pim}{BR_f1_2pip2pim}{}{0.1100}{0.1100}{0}%
\htquantdef{BR_f1_2pizpippim}{BR_f1_2pizpippim}{}{0.2200}{0.2200}{0}%
\htquantdef{BR_KS_2piz}{BR_KS_2piz}{}{0.3069}{0.3069}{0}%
\htquantdef{BR_KS_pimpip}{BR_KS_pimpip}{}{0.6920}{0.6920}{0}%
\htquantdef{BR_om_pimpip}{BR_om_pimpip}{}{1.530\cdot 10^{-2}}{1.530\cdot 10^{-2}}{0}%
\htquantdef{BR_om_pimpippiz}{BR_om_pimpippiz}{}{0.8920}{0.8920}{0}%
\htquantdef{BR_om_pizgamma}{BR_om_pizgamma}{}{8.280\cdot 10^{-2}}{8.280\cdot 10^{-2}}{0}%
\htquantdef{BR_phi_KmKp}{BR_phi_KmKp}{}{0.4890}{0.4890}{0}%
\htquantdef{BR_phi_KSKL}{BR_phi_KSKL}{}{0.3420}{0.3420}{0}%
\htquantdef{BRA_Kz_KL_KET}{BRA_Kz_KL_KET}{}{0.5000}{0.5000}{0}%
\htquantdef{BRA_Kz_KS_KET}{BRA_Kz_KS_KET}{}{0.5000}{0.5000}{0}%
\htquantdef{BRA_Kzbar_KL_KET}{BRA_Kzbar_KL_KET}{}{0.5000}{0.5000}{0}%
\htquantdef{BRA_Kzbar_KS_KET}{BRA_Kzbar_KS_KET}{}{0.5000}{0.5000}{0}%
\htquantdef{delta_K}{delta_K}{}{\ensuremath{(0.9000 \pm 0.2200) \cdot 10^{-2}}}{0.9000\cdot 10^{-2}}{0.2200\cdot 10^{-2}}%
\htquantdef{delta_LD_tauK_Kmu}{delta_LD_tauK_Kmu}{}{0.90 \pm 0.22}{0.90}{0.22}%
\htquantdef{delta_LD_taupi_pimu}{delta_LD_taupi_pimu}{}{0.16 \pm 0.14}{0.16}{0.14}%
\htquantdef{delta_mu_gamma}{delta_mu_gamma}{}{0.9958}{0.9958}{0}%
\htquantdef{delta_mu_W}{delta_mu_W}{}{\ensuremath{1.00000103622 \pm 0.00000000039}}{1.00000103622}{0.00000000039}%
\htquantdef{delta_pi}{delta_pi}{}{\ensuremath{(0.1600 \pm 0.1400) \cdot 10^{-2}}}{0.1600\cdot 10^{-2}}{0.1400\cdot 10^{-2}}%
\htquantdef{delta_tau_gamma}{delta_tau_gamma}{}{0.9957}{0.9957}{0}%
\htquantdef{delta_tau_W}{delta_tau_W}{}{\ensuremath{1.00029304 \pm 0.00000012}}{1.00029304}{0.00000012}%
\htquantdef{deltaR_su3break}{deltaR_su3break}{}{0.239 \pm 0.030}{0.239}{0.030}%
\htquantdef{deltaR_su3break_d2pert}{deltaR_su3break_d2pert}{}{\ensuremath{9.300 \pm 3.400}}{9.300}{3.400}%
\htquantdef{deltaR_su3break_pheno}{deltaR_su3break_pheno}{}{\ensuremath{0.1544 \pm 0.0037}}{0.1544}{0.0037}%
\htquantdef{deltaR_su3break_remain}{deltaR_su3break_remain}{}{\ensuremath{(0.3400 \pm 0.2800) \cdot 10^{-2}}}{0.3400\cdot 10^{-2}}{0.2800\cdot 10^{-2}}%
\htquantdef{f_K}{f_K}{}{156.3 \pm 0.9}{156.3}{0.9}%
\htquantdef{f_K_by_f_pi}{f_K_by_f_pi}{}{\ensuremath{1.1920 \pm 0.0050}}{1.1920}{0.0050}%
\htquantdef{fp0_Kpi}{fp0_Kpi}{}{\ensuremath{0.9661 \pm 0.0032}}{0.9661}{0.0032}%
\htquantdef{G_F_by_hcut3_c3}{G_F_by_hcut3_c3}{}{\ensuremath{(1.16637870 \pm 0.00000060) \cdot 10^{-11}}}{1.16637870\cdot 10^{-11}}{0.00000060\cdot 10^{-11}}%
\htquantdef{Gamma10}{\Gamma_{10}}{\BRF{\tau^-}{K^- \nu_\tau}}{\ensuremath{(0.6955 \pm 0.0096) \cdot 10^{-2}}}{0.6955\cdot 10^{-2}}{0.0096\cdot 10^{-2}}%
\htquantdef{Gamma102}{\Gamma_{102}}{\BRF{\tau^-}{3h^- 2h^+ \ge{}0 \text{neutrals} \nu_\tau\;(\text{ex.} K^0)}}{\ensuremath{(9.632 \pm 0.275) \cdot 10^{-4}}}{9.632\cdot 10^{-4}}{0.275\cdot 10^{-4}}%
\htquantdef{Gamma103}{\Gamma_{103}}{\BRF{\tau^-}{3h^- 2h^+ \nu_\tau ~(\text{ex.~}K^0)}}{\ensuremath{(8.342 \pm 0.242) \cdot 10^{-4}}}{8.342\cdot 10^{-4}}{0.242\cdot 10^{-4}}%
\htquantdef{Gamma104}{\Gamma_{104}}{\BRF{\tau^-}{3h^- 2h^+ \pi^0 \nu_\tau ~(\text{ex.~}K^0)}}{\ensuremath{(1.290 \pm 0.068) \cdot 10^{-4}}}{1.290\cdot 10^{-4}}{0.068\cdot 10^{-4}}%
\htquantdef{Gamma10by5}{\frac{\Gamma_{10}}{\Gamma_{5}}}{\frac{\BRF{\tau^-}{K^- \nu_\tau}}{\BRF{\tau^-}{e^- \bar{\nu}_e \nu_\tau}}}{\ensuremath{(3.903 \pm 0.054) \cdot 10^{-2}}}{3.903\cdot 10^{-2}}{0.054\cdot 10^{-2}}%
\htquantdef{Gamma10by9}{\frac{\Gamma_{10}}{\Gamma_{9}}}{\frac{\BRF{\tau^-}{K^- \nu_\tau}}{\BRF{\tau^-}{\pi^- \nu_\tau}}}{\ensuremath{(6.431 \pm 0.094) \cdot 10^{-2}}}{6.431\cdot 10^{-2}}{0.094\cdot 10^{-2}}%
\htquantdef{Gamma110}{\Gamma_{110}}{\BRF{\tau^-}{X_s^- \nu_\tau}}{\ensuremath{(2.882 \pm 0.047) \cdot 10^{-2}}}{2.882\cdot 10^{-2}}{0.047\cdot 10^{-2}}%
\htquantdef{Gamma110_pdg09}{\Gamma_{110}_pdg09}{}{\ensuremath{(2.825 \pm 0.036) \cdot 10^{-2}}}{2.825\cdot 10^{-2}}{0.036\cdot 10^{-2}}%
\htquantdef{Gamma126}{\Gamma_{126}}{\BRF{\tau^-}{\pi^- \pi^0 \eta \nu_\tau}}{\ensuremath{(0.1386 \pm 0.0072) \cdot 10^{-2}}}{0.1386\cdot 10^{-2}}{0.0072\cdot 10^{-2}}%
\htquantdef{Gamma128}{\Gamma_{128}}{\BRF{\tau^-}{K^- \eta \nu_\tau}}{\ensuremath{(1.530 \pm 0.081) \cdot 10^{-4}}}{1.530\cdot 10^{-4}}{0.081\cdot 10^{-4}}%
\htquantdef{Gamma13}{\Gamma_{13}}{\BRF{\tau^-}{h^- \pi^0 \nu_\tau}}{\ensuremath{0.25937 \pm 0.00090}}{0.25937}{0.00090}%
\htquantdef{Gamma130}{\Gamma_{130}}{\BRF{\tau^-}{K^- \pi^0 \eta \nu_\tau}}{\ensuremath{(4.827 \pm 1.161) \cdot 10^{-5}}}{4.827\cdot 10^{-5}}{1.161\cdot 10^{-5}}%
\htquantdef{Gamma132}{\Gamma_{132}}{\BRF{\tau^-}{\pi^- \bar{K}^0 \eta \nu_\tau}}{\ensuremath{(9.332 \pm 1.490) \cdot 10^{-5}}}{9.332\cdot 10^{-5}}{1.490\cdot 10^{-5}}%
\htquantdef{Gamma136}{\Gamma_{136}}{\BRF{\tau^-}{\pi^- \pi^- \pi^+ \eta \nu_\tau\;(\text{ex.} K^0)}}{\ensuremath{(2.224 \pm 0.119) \cdot 10^{-4}}}{2.224\cdot 10^{-4}}{0.119\cdot 10^{-4}}%
\htquantdef{Gamma14}{\Gamma_{14}}{\BRF{\tau^-}{\pi^- \pi^0 \nu_\tau}}{\ensuremath{0.25504 \pm 0.00092}}{0.25504}{0.00092}%
\htquantdef{Gamma150}{\Gamma_{150}}{\BRF{\tau^-}{h^- \omega \nu_\tau}}{\ensuremath{(1.995 \pm 0.064) \cdot 10^{-2}}}{1.995\cdot 10^{-2}}{0.064\cdot 10^{-2}}%
\htquantdef{Gamma150by66}{\frac{\Gamma_{150}}{\Gamma_{66}}}{\frac{\BRF{\tau^-}{h^- \omega \nu_\tau}}{\BRF{\tau^-}{h^- h^- h^+ \pi^0 \nu_\tau\;(\text{ex.} K^0)}}}{\ensuremath{0.4333 \pm 0.0139}}{0.4333}{0.0139}%
\htquantdef{Gamma151}{\Gamma_{151}}{\BRF{\tau^-}{K^- \omega \nu_\tau}}{\ensuremath{(4.100 \pm 0.922) \cdot 10^{-4}}}{4.100\cdot 10^{-4}}{0.922\cdot 10^{-4}}%
\htquantdef{Gamma152}{\Gamma_{152}}{\BRF{\tau^-}{h^- \pi^0 \omega \nu_\tau}}{\ensuremath{(0.4051 \pm 0.0418) \cdot 10^{-2}}}{0.4051\cdot 10^{-2}}{0.0418\cdot 10^{-2}}%
\htquantdef{Gamma152by76}{\frac{\Gamma_{152}}{\Gamma_{76}}}{\frac{\BRF{\tau^-}{h^- \omega \pi^0 \nu_\tau}}{\BRF{\tau^-}{h^- h^- h^+ 2\pi^0 \nu_\tau\;(\text{ex.} K^0)}}}{\ensuremath{0.8245 \pm 0.0757}}{0.8245}{0.0757}%
\htquantdef{Gamma16}{\Gamma_{16}}{\BRF{\tau^-}{K^- \pi^0 \nu_\tau}}{\ensuremath{(0.4331 \pm 0.0149) \cdot 10^{-2}}}{0.4331\cdot 10^{-2}}{0.0149\cdot 10^{-2}}%
\htquantdef{Gamma17}{\Gamma_{17}}{\BRF{\tau^-}{h^- \ge{}2 \pi^0 \nu_\tau}}{\ensuremath{0.10804 \pm 0.00095}}{0.10804}{0.00095}%
\htquantdef{Gamma19}{\Gamma_{19}}{\BRF{\tau^-}{h^- 2\pi^0 \nu_\tau\;(\text{ex.} K^0)}}{\ensuremath{(9.302 \pm 0.097) \cdot 10^{-2}}}{9.302\cdot 10^{-2}}{0.097\cdot 10^{-2}}%
\htquantdef{Gamma19by13}{\frac{\Gamma_{19}}{\Gamma_{13}}}{\frac{\BRF{\tau^-}{h^- 2\pi^0 \nu_\tau\;(\text{ex.} K^0)}}{\BRF{\tau^-}{h^- \pi^0 \nu_\tau}}}{\ensuremath{0.3586 \pm 0.0044}}{0.3586}{0.0044}%
\htquantdef{Gamma20}{\Gamma_{20}}{\BRF{\tau^-}{\pi^- 2\pi^0 \nu_\tau ~(\text{ex.~}K^0)}}{\ensuremath{(9.239 \pm 0.100) \cdot 10^{-2}}}{9.239\cdot 10^{-2}}{0.100\cdot 10^{-2}}%
\htquantdef{Gamma23}{\Gamma_{23}}{\BRF{\tau^-}{K^- 2\pi^0 \nu_\tau ~(\text{ex.~}K^0)}}{\ensuremath{(6.303 \pm 2.204) \cdot 10^{-4}}}{6.303\cdot 10^{-4}}{2.204\cdot 10^{-4}}%
\htquantdef{Gamma25}{\Gamma_{25}}{\BRF{\tau^-}{h^- \ge{} 3\pi^0 \nu_\tau\;(\text{ex.} K^0)}}{\ensuremath{(1.234 \pm 0.065) \cdot 10^{-2}}}{1.234\cdot 10^{-2}}{0.065\cdot 10^{-2}}%
\htquantdef{Gamma26}{\Gamma_{26}}{\BRF{\tau^-}{h^- 3\pi^0 \nu_\tau}}{\ensuremath{(1.157 \pm 0.072) \cdot 10^{-2}}}{1.157\cdot 10^{-2}}{0.072\cdot 10^{-2}}%
\htquantdef{Gamma26by13}{\frac{\Gamma_{26}}{\Gamma_{13}}}{\frac{\BRF{\tau^-}{h^- 3\pi^0 \nu_\tau}}{\BRF{\tau^-}{h^- \pi^0 \nu_\tau}}}{\ensuremath{(4.459 \pm 0.277) \cdot 10^{-2}}}{4.459\cdot 10^{-2}}{0.277\cdot 10^{-2}}%
\htquantdef{Gamma27}{\Gamma_{27}}{\BRF{\tau^-}{\pi^- 3\pi^0 \nu_\tau ~(\text{ex.~}K^0)}}{\ensuremath{(1.030 \pm 0.075) \cdot 10^{-2}}}{1.030\cdot 10^{-2}}{0.075\cdot 10^{-2}}%
\htquantdef{Gamma28}{\Gamma_{28}}{\BRF{\tau^-}{K^- 3\pi^0 \nu_\tau ~(\text{ex.~}K^0,\eta)}}{\ensuremath{(4.193 \pm 2.160) \cdot 10^{-4}}}{4.193\cdot 10^{-4}}{2.160\cdot 10^{-4}}%
\htquantdef{Gamma29}{\Gamma_{29}}{\BRF{\tau^-}{h^- 4\pi^0 \nu_\tau\;(\text{ex.} K^0)}}{\ensuremath{(0.1572 \pm 0.0391) \cdot 10^{-2}}}{0.1572\cdot 10^{-2}}{0.0391\cdot 10^{-2}}%
\htquantdef{Gamma3}{\Gamma_{3}}{\BRF{\tau^-}{\mu^- \bar{\nu}_\mu \nu_\tau}}{\ensuremath{0.17392 \pm 0.00040}}{0.17392}{0.00040}%
\htquantdef{Gamma30}{\Gamma_{30}}{\BRF{\tau^-}{h^- 4\pi^0 \nu_\tau ~(\text{ex.~}K^0,\eta)}}{\ensuremath{(0.1103 \pm 0.0391) \cdot 10^{-2}}}{0.1103\cdot 10^{-2}}{0.0391\cdot 10^{-2}}%
\htquantdef{Gamma31}{\Gamma_{31}}{\BRF{\tau^-}{K^- \ge{}0 \pi^0 \ge{}0 K^0 \ge{}0 \gamma \nu_\tau}}{\ensuremath{(1.548 \pm 0.030) \cdot 10^{-2}}}{1.548\cdot 10^{-2}}{0.030\cdot 10^{-2}}%
\htquantdef{Gamma33}{\Gamma_{33}}{\BRF{\tau^-}{K_S^0 (\text{particles})^- \nu_\tau}}{\ensuremath{(0.9019 \pm 0.0081) \cdot 10^{-2}}}{0.9019\cdot 10^{-2}}{0.0081\cdot 10^{-2}}%
\htquantdef{Gamma34}{\Gamma_{34}}{\BRF{\tau^-}{h^- \bar{K}^0 \nu_\tau}}{\ensuremath{(0.9878 \pm 0.0119) \cdot 10^{-2}}}{0.9878\cdot 10^{-2}}{0.0119\cdot 10^{-2}}%
\htquantdef{Gamma35}{\Gamma_{35}}{\BRF{\tau^-}{\pi^- \bar{K}^0 \nu_\tau}}{\ensuremath{(0.8378 \pm 0.0123) \cdot 10^{-2}}}{0.8378\cdot 10^{-2}}{0.0123\cdot 10^{-2}}%
\htquantdef{Gamma37}{\Gamma_{37}}{\BRF{\tau^-}{K^- K^0 \nu_\tau}}{\ensuremath{(0.1500 \pm 0.0050) \cdot 10^{-2}}}{0.1500\cdot 10^{-2}}{0.0050\cdot 10^{-2}}%
\htquantdef{Gamma38}{\Gamma_{38}}{\BRF{\tau^-}{K^- K^0 \ge{}0 \pi^0 \nu_\tau}}{\ensuremath{(0.3029 \pm 0.0074) \cdot 10^{-2}}}{0.3029\cdot 10^{-2}}{0.0074\cdot 10^{-2}}%
\htquantdef{Gamma39}{\Gamma_{39}}{\BRF{\tau^-}{h^- \bar{K}^0 \pi^0 \nu_\tau}}{\ensuremath{(0.5208 \pm 0.0114) \cdot 10^{-2}}}{0.5208\cdot 10^{-2}}{0.0114\cdot 10^{-2}}%
\htquantdef{Gamma3by5}{\frac{\Gamma_{3}}{\Gamma_{5}}}{\frac{\BRF{\tau^-}{\mu^- \bar{\nu}_\mu \nu_\tau}}{\BRF{\tau^-}{e^- \bar{\nu}_e \nu_\tau}}}{\ensuremath{0.9761 \pm 0.0028}}{0.9761}{0.0028}%
\htquantdef{Gamma40}{\Gamma_{40}}{\BRF{\tau^-}{\pi^- \bar{K}^0 \pi^0 \nu_\tau}}{\ensuremath{(0.3680 \pm 0.0103) \cdot 10^{-2}}}{0.3680\cdot 10^{-2}}{0.0103\cdot 10^{-2}}%
\htquantdef{Gamma42}{\Gamma_{42}}{\BRF{\tau^-}{K^- \pi^0 K^0 \nu_\tau}}{\ensuremath{(0.1528 \pm 0.0070) \cdot 10^{-2}}}{0.1528\cdot 10^{-2}}{0.0070\cdot 10^{-2}}%
\htquantdef{Gamma43}{\Gamma_{43}}{\BRF{\tau^-}{\pi^- \bar{K}^0 \ge{}1 \pi^0 \nu_\tau}}{\ensuremath{(0.3805 \pm 0.0229) \cdot 10^{-2}}}{0.3805\cdot 10^{-2}}{0.0229\cdot 10^{-2}}%
\htquantdef{Gamma44}{\Gamma_{44}}{\BRF{\tau^-}{\pi^- \bar{K}^0 \pi^0 \pi^0 \nu_\tau ~(\text{ex.~}K^0)}}{\ensuremath{(1.245 \pm 2.043) \cdot 10^{-4}}}{1.245\cdot 10^{-4}}{2.043\cdot 10^{-4}}%
\htquantdef{Gamma46}{\Gamma_{46}}{\BRF{\tau^-}{\pi^- K^0 \bar{K}^0 \nu_\tau}}{\ensuremath{(0.1329 \pm 0.0110) \cdot 10^{-2}}}{0.1329\cdot 10^{-2}}{0.0110\cdot 10^{-2}}%
\htquantdef{Gamma47}{\Gamma_{47}}{\BRF{\tau^-}{\pi^- K_S^0 K_S^0 \nu_\tau}}{\ensuremath{(2.359 \pm 0.061) \cdot 10^{-4}}}{2.359\cdot 10^{-4}}{0.061\cdot 10^{-4}}%
\htquantdef{Gamma48}{\Gamma_{48}}{\BRF{\tau^-}{\pi^- K_S^0 K_L^0 \nu_\tau}}{\ensuremath{(8.574 \pm 1.037) \cdot 10^{-4}}}{8.574\cdot 10^{-4}}{1.037\cdot 10^{-4}}%
\htquantdef{Gamma49}{\Gamma_{49}}{\BRF{\tau^-}{\pi^- K^0 \bar{K}^0 \pi^0 \nu_\tau}}{\ensuremath{(2.896 \pm 1.051) \cdot 10^{-4}}}{2.896\cdot 10^{-4}}{1.051\cdot 10^{-4}}%
\htquantdef{Gamma5}{\Gamma_{5}}{\BRF{\tau^-}{e^- \bar{\nu}_e \nu_\tau}}{\ensuremath{0.17817 \pm 0.00041}}{0.17817}{0.00041}%
\htquantdef{Gamma50}{\Gamma_{50}}{\BRF{\tau^-}{\pi^- \pi^0 K_S^0 K_S^0 \nu_\tau}}{\ensuremath{(1.844 \pm 0.206) \cdot 10^{-5}}}{1.844\cdot 10^{-5}}{0.206\cdot 10^{-5}}%
\htquantdef{Gamma51}{\Gamma_{51}}{\BRF{\tau^-}{\pi^- \pi^0 K_S^0 K_L^0 \nu_\tau}}{\ensuremath{(2.527 \pm 1.047) \cdot 10^{-4}}}{2.527\cdot 10^{-4}}{1.047\cdot 10^{-4}}%
\htquantdef{Gamma53}{\Gamma_{53}}{\BRF{\tau^-}{\bar{K}^0 h^- h^- h^+ \nu_\tau}}{\ensuremath{(2.221 \pm 2.024) \cdot 10^{-4}}}{2.221\cdot 10^{-4}}{2.024\cdot 10^{-4}}%
\htquantdef{Gamma54}{\Gamma_{54}}{\BRF{\tau^-}{h^- h^- h^+ \ge{}0 \text{neutrals} \ge{}0 K_L^0 \nu_\tau}}{\ensuremath{0.15201 \pm 0.00059}}{0.15201}{0.00059}%
\htquantdef{Gamma55}{\Gamma_{55}}{\BRF{\tau^-}{h^- h^- h^+ \ge{}0 \text{neutrals} \nu_\tau\;(\text{ex.} K^0)}}{\ensuremath{0.14573 \pm 0.00056}}{0.14573}{0.00056}%
\htquantdef{Gamma57}{\Gamma_{57}}{\BRF{\tau^-}{h^- h^- h^+ \nu_\tau\;(\text{ex.} K^0)}}{\ensuremath{(9.449 \pm 0.053) \cdot 10^{-2}}}{9.449\cdot 10^{-2}}{0.053\cdot 10^{-2}}%
\htquantdef{Gamma57by55}{\frac{\Gamma_{57}}{\Gamma_{55}}}{\frac{\BRF{\tau^-}{h^- h^- h^+ \nu_\tau\;(\text{ex.} K^0)}}{\BRF{\tau^-}{h^- h^- h^+ \ge{}0 \text{neutrals} \nu_\tau\;(\text{ex.} K^0)}}}{\ensuremath{0.6484 \pm 0.0029}}{0.6484}{0.0029}%
\htquantdef{Gamma58}{\Gamma_{58}}{\BRF{\tau^-}{h^- h^- h^+ \nu_\tau\;(\text{ex.} K^0, \omega)}}{\ensuremath{(9.418 \pm 0.053) \cdot 10^{-2}}}{9.418\cdot 10^{-2}}{0.053\cdot 10^{-2}}%
\htquantdef{Gamma60}{\Gamma_{60}}{\BRF{\tau^-}{\pi^- \pi^- \pi^+ \nu_\tau\;(\text{ex.} K^0)}}{\ensuremath{(9.010 \pm 0.051) \cdot 10^{-2}}}{9.010\cdot 10^{-2}}{0.051\cdot 10^{-2}}%
\htquantdef{Gamma62}{\Gamma_{62}}{\BRF{\tau^-}{\pi^- \pi^- \pi^+ \nu_\tau ~(\text{ex.~}K^0,\omega)}}{\ensuremath{(8.980 \pm 0.051) \cdot 10^{-2}}}{8.980\cdot 10^{-2}}{0.051\cdot 10^{-2}}%
\htquantdef{Gamma66}{\Gamma_{66}}{\BRF{\tau^-}{h^- h^- h^+ \pi^0 \nu_\tau\;(\text{ex.} K^0)}}{\ensuremath{(4.604 \pm 0.051) \cdot 10^{-2}}}{4.604\cdot 10^{-2}}{0.051\cdot 10^{-2}}%
\htquantdef{Gamma69}{\Gamma_{69}}{\BRF{\tau^-}{\pi^- \pi^- \pi^+ \pi^0 \nu_\tau\;(\text{ex.} K^0)}}{\ensuremath{(4.516 \pm 0.052) \cdot 10^{-2}}}{4.516\cdot 10^{-2}}{0.052\cdot 10^{-2}}%
\htquantdef{Gamma7}{\Gamma_{7}}{\BRF{\tau^-}{h^- \ge{}0 K_L^0 \nu_\tau}}{\ensuremath{0.12027 \pm 0.00054}}{0.12027}{0.00054}%
\htquantdef{Gamma70}{\Gamma_{70}}{\BRF{\tau^-}{\pi^- \pi^- \pi^+ \pi^0 \nu_\tau ~(\text{ex.~}K^0,\omega)}}{\ensuremath{(2.767 \pm 0.071) \cdot 10^{-2}}}{2.767\cdot 10^{-2}}{0.071\cdot 10^{-2}}%
\htquantdef{Gamma74}{\Gamma_{74}}{\BRF{\tau^-}{h^- h^- h^+ \ge{} 2\pi^0 \nu_\tau\;(\text{ex.} K^0)}}{\ensuremath{(0.5128 \pm 0.0311) \cdot 10^{-2}}}{0.5128\cdot 10^{-2}}{0.0311\cdot 10^{-2}}%
\htquantdef{Gamma76}{\Gamma_{76}}{\BRF{\tau^-}{h^- h^- h^+ 2\pi^0 \nu_\tau\;(\text{ex.} K^0)}}{\ensuremath{(0.4913 \pm 0.0310) \cdot 10^{-2}}}{0.4913\cdot 10^{-2}}{0.0310\cdot 10^{-2}}%
\htquantdef{Gamma76by54}{\frac{\Gamma_{76}}{\Gamma_{54}}}{\frac{\BRF{\tau^-}{h^- h^- h^+ 2\pi^0 \nu_\tau\;(\text{ex.} K^0)}}{\BRF{\tau^-}{h^- h^- h^+ \ge{}0 \text{neutrals} \ge{}0 K_L^0 \nu_\tau}}}{\ensuremath{(3.232 \pm 0.202) \cdot 10^{-2}}}{3.232\cdot 10^{-2}}{0.202\cdot 10^{-2}}%
\htquantdef{Gamma77}{\Gamma_{77}}{\BRF{\tau^-}{h^- h^- h^+ 2\pi^0 \nu_\tau ~(\text{ex.~}K^0,\omega,\eta)}}{\ensuremath{(9.711 \pm 3.543) \cdot 10^{-4}}}{9.711\cdot 10^{-4}}{3.543\cdot 10^{-4}}%
\htquantdef{Gamma78}{\Gamma_{78}}{\BRF{\tau^-}{h^- h^- h^+ 3\pi^0 \nu_\tau}}{\ensuremath{(2.153 \pm 0.298) \cdot 10^{-4}}}{2.153\cdot 10^{-4}}{0.298\cdot 10^{-4}}%
\htquantdef{Gamma8}{\Gamma_{8}}{\BRF{\tau^-}{h^- \nu_\tau}}{\ensuremath{0.11509 \pm 0.00054}}{0.11509}{0.00054}%
\htquantdef{Gamma800}{\Gamma_{800}}{\BRF{\tau^-}{\pi^- \omega \nu_\tau}}{\ensuremath{(1.954 \pm 0.065) \cdot 10^{-2}}}{1.954\cdot 10^{-2}}{0.065\cdot 10^{-2}}%
\htquantdef{Gamma801}{\Gamma_{801}}{\BRF{\tau^-}{K^- \phi \nu_\tau (\phi \to KK)}}{\ensuremath{(3.664 \pm 1.360) \cdot 10^{-5}}}{3.664\cdot 10^{-5}}{1.360\cdot 10^{-5}}%
\htquantdef{Gamma802}{\Gamma_{802}}{\BRF{\tau^-}{K^- \pi^- \pi^+ \nu_\tau ~(\text{ex.~}K^0,\omega)}}{\ensuremath{(0.2923 \pm 0.0068) \cdot 10^{-2}}}{0.2923\cdot 10^{-2}}{0.0068\cdot 10^{-2}}%
\htquantdef{Gamma803}{\Gamma_{803}}{\BRF{\tau^-}{K^- \pi^- \pi^+ \pi^0 \nu_\tau ~(\text{ex.~}K^0,\omega,\eta)}}{\ensuremath{(4.105 \pm 1.429) \cdot 10^{-4}}}{4.105\cdot 10^{-4}}{1.429\cdot 10^{-4}}%
\htquantdef{Gamma804}{\Gamma_{804}}{\BRF{\tau^-}{\pi^- K_L^0 K_L^0 \nu_\tau}}{\ensuremath{(2.359 \pm 0.061) \cdot 10^{-4}}}{2.359\cdot 10^{-4}}{0.061\cdot 10^{-4}}%
\htquantdef{Gamma805}{\Gamma_{805}}{\BRF{\tau^-}{a_1^- (\to \pi^- \gamma) \nu_\tau}}{\ensuremath{(4.000 \pm 2.000) \cdot 10^{-4}}}{4.000\cdot 10^{-4}}{2.000\cdot 10^{-4}}%
\htquantdef{Gamma806}{\Gamma_{806}}{\BRF{\tau^-}{\pi^- \pi^0 K_L^0 K_L^0 \nu_\tau}}{\ensuremath{(1.844 \pm 0.206) \cdot 10^{-5}}}{1.844\cdot 10^{-5}}{0.206\cdot 10^{-5}}%
\htquantdef{Gamma80by60}{\frac{\Gamma_{80}}{\Gamma_{60}}}{\frac{\BRF{\tau^-}{K^- \pi^- h^+ \nu_\tau\;(\text{ex.} K^0)}}{\BRF{\tau^-}{\pi^- \pi^- \pi^+ \nu_\tau\;(\text{ex.} K^0)}}}{\ensuremath{(4.845 \pm 0.081) \cdot 10^{-2}}}{4.845\cdot 10^{-2}}{0.081\cdot 10^{-2}}%
\htquantdef{Gamma810}{\Gamma_{810}}{\BRF{\tau^-}{2\pi^- \pi^+ 3\pi^0 \nu_\tau ~(\text{ex.~}K^0)}}{\ensuremath{(1.969 \pm 0.298) \cdot 10^{-4}}}{1.969\cdot 10^{-4}}{0.298\cdot 10^{-4}}%
\htquantdef{Gamma811}{\Gamma_{811}}{\BRF{\tau^-}{\pi^- 2\pi^0 \omega \nu_\tau ~(\text{ex.~}K^0)}}{\ensuremath{(7.143 \pm 1.568) \cdot 10^{-5}}}{7.143\cdot 10^{-5}}{1.568\cdot 10^{-5}}%
\htquantdef{Gamma812}{\Gamma_{812}}{\BRF{\tau^-}{2\pi^- \pi^+ 3\pi^0 \nu_\tau ~(\text{ex.~}K^0, \eta, \omega, f_1)}}{\ensuremath{(1.166 \pm 2.683) \cdot 10^{-5}}}{1.166\cdot 10^{-5}}{2.683\cdot 10^{-5}}%
\htquantdef{Gamma81by69}{\frac{\Gamma_{81}}{\Gamma_{69}}}{\frac{\BRF{\tau^-}{K^- \pi^- h^+ \pi^0 \nu_\tau\;(\text{ex.} K^0)}}{\BRF{\tau^-}{\pi^- \pi^- \pi^+ \pi^0 \nu_\tau\;(\text{ex.} K^0)}}}{\ensuremath{(1.932 \pm 0.266) \cdot 10^{-2}}}{1.932\cdot 10^{-2}}{0.266\cdot 10^{-2}}%
\htquantdef{Gamma82}{\Gamma_{82}}{\BRF{\tau^-}{K^- \pi^- \pi^+ \ge{}0 \text{neutrals} \nu_\tau}}{\ensuremath{(0.4796 \pm 0.0138) \cdot 10^{-2}}}{0.4796\cdot 10^{-2}}{0.0138\cdot 10^{-2}}%
\htquantdef{Gamma820}{\Gamma_{820}}{\BRF{\tau^-}{3\pi^- 2\pi^+ \nu_\tau ~(\text{ex.~}K^0, \omega)}}{\ensuremath{(8.323 \pm 0.242) \cdot 10^{-4}}}{8.323\cdot 10^{-4}}{0.242\cdot 10^{-4}}%
\htquantdef{Gamma821}{\Gamma_{821}}{\BRF{\tau^-}{3\pi^- 2\pi^+ \nu_\tau ~(\text{ex.~}K^0, \omega, f_1)}}{\ensuremath{(7.795 \pm 0.229) \cdot 10^{-4}}}{7.795\cdot 10^{-4}}{0.229\cdot 10^{-4}}%
\htquantdef{Gamma822}{\Gamma_{822}}{\BRF{\tau^-}{K^- 2\pi^- 2\pi^+ \nu_\tau ~(\text{ex.~}K^0)}}{\ensuremath{(0.599 \pm 1.208) \cdot 10^{-6}}}{0.599\cdot 10^{-6}}{1.208\cdot 10^{-6}}%
\htquantdef{Gamma830}{\Gamma_{830}}{\BRF{\tau^-}{3\pi^- 2\pi^+ \pi^0 \nu_\tau ~(\text{ex.~}K^0)}}{\ensuremath{(1.639 \pm 0.121) \cdot 10^{-4}}}{1.639\cdot 10^{-4}}{0.121\cdot 10^{-4}}%
\htquantdef{Gamma831}{\Gamma_{831}}{\BRF{\tau^-}{2\pi^- \pi^+ \omega \nu_\tau ~(\text{ex.~}K^0)}}{\ensuremath{(8.618 \pm 0.601) \cdot 10^{-5}}}{8.618\cdot 10^{-5}}{0.601\cdot 10^{-5}}%
\htquantdef{Gamma832}{\Gamma_{832}}{\BRF{\tau^-}{3\pi^- 2\pi^+ \pi^0 \nu_\tau ~(\text{ex.~}K^0, \eta, \omega, f_1)}}{\ensuremath{(3.607 \pm 0.936) \cdot 10^{-5}}}{3.607\cdot 10^{-5}}{0.936\cdot 10^{-5}}%
\htquantdef{Gamma833}{\Gamma_{833}}{\BRF{\tau^-}{K^- 2\pi^- 2\pi^+ \pi^0 \nu_\tau ~(\text{ex.~}K^0)}}{\ensuremath{(1.125 \pm 0.566) \cdot 10^{-6}}}{1.125\cdot 10^{-6}}{0.566\cdot 10^{-6}}%
\htquantdef{Gamma85}{\Gamma_{85}}{\BRF{\tau^-}{K^- \pi^- \pi^+ \nu_\tau\;(\text{ex.} K^0)}}{\ensuremath{(0.2929 \pm 0.0068) \cdot 10^{-2}}}{0.2929\cdot 10^{-2}}{0.0068\cdot 10^{-2}}%
\htquantdef{Gamma88}{\Gamma_{88}}{\BRF{\tau^-}{K^- \pi^- \pi^+ \pi^0 \nu_\tau\;(\text{ex.} K^0)}}{\ensuremath{(8.113 \pm 1.168) \cdot 10^{-4}}}{8.113\cdot 10^{-4}}{1.168\cdot 10^{-4}}%
\htquantdef{Gamma89}{\Gamma_{89}}{\BRF{\tau^-}{K^- \pi^- \pi^+ \pi^0 \nu_\tau\;(\text{ex.} K^0, \eta)}}{\ensuremath{(7.762 \pm 1.168) \cdot 10^{-4}}}{7.762\cdot 10^{-4}}{1.168\cdot 10^{-4}}%
\htquantdef{Gamma9}{\Gamma_{9}}{\BRF{\tau^-}{\pi^- \nu_\tau}}{\ensuremath{0.10814 \pm 0.00053}}{0.10814}{0.00053}%
\htquantdef{Gamma910}{\Gamma_{910}}{\BRF{\tau^-}{2\pi^- \pi^+ \eta \nu_\tau (\eta \to 3\pi^0) ~(\text{ex.~}K^0)}}{\ensuremath{(7.267 \pm 0.389) \cdot 10^{-5}}}{7.267\cdot 10^{-5}}{0.389\cdot 10^{-5}}%
\htquantdef{Gamma911}{\Gamma_{911}}{\BRF{\tau^-}{\pi^- 2\pi^0 \eta \nu_\tau (\eta \to \pi^+ \pi^- \pi^0) ~(\text{ex.~}K^0)}}{\ensuremath{(4.118 \pm 0.690) \cdot 10^{-5}}}{4.118\cdot 10^{-5}}{0.690\cdot 10^{-5}}%
\htquantdef{Gamma92}{\Gamma_{92}}{\BRF{\tau^-}{\pi^- K^- K^+ \ge{}0 \text{neutrals} \nu_\tau}}{\ensuremath{(0.1497 \pm 0.0033) \cdot 10^{-2}}}{0.1497\cdot 10^{-2}}{0.0033\cdot 10^{-2}}%
\htquantdef{Gamma920}{\Gamma_{920}}{\BRF{\tau^-}{\pi^- f_1 \nu_\tau (f_1 \to 2\pi^- 2\pi^+)}}{\ensuremath{(5.278 \pm 0.424) \cdot 10^{-5}}}{5.278\cdot 10^{-5}}{0.424\cdot 10^{-5}}%
\htquantdef{Gamma93}{\Gamma_{93}}{\BRF{\tau^-}{\pi^- K^- K^+ \nu_\tau}}{\ensuremath{(0.1436 \pm 0.0027) \cdot 10^{-2}}}{0.1436\cdot 10^{-2}}{0.0027\cdot 10^{-2}}%
\htquantdef{Gamma930}{\Gamma_{930}}{\BRF{\tau^-}{2\pi^- \pi^+ \eta \nu_\tau (\eta \to \pi^+\pi^-\pi^0) ~(\text{ex.~}K^0)}}{\ensuremath{(5.097 \pm 0.273) \cdot 10^{-5}}}{5.097\cdot 10^{-5}}{0.273\cdot 10^{-5}}%
\htquantdef{Gamma93by60}{\frac{\Gamma_{93}}{\Gamma_{60}}}{\frac{\BRF{\tau^-}{\pi^- K^- K^+ \nu_\tau}}{\BRF{\tau^-}{\pi^- \pi^- \pi^+ \nu_\tau\;(\text{ex.} K^0)}}}{\ensuremath{(1.594 \pm 0.030) \cdot 10^{-2}}}{1.594\cdot 10^{-2}}{0.030\cdot 10^{-2}}%
\htquantdef{Gamma94}{\Gamma_{94}}{\BRF{\tau^-}{\pi^- K^- K^+ \pi^0 \nu_\tau}}{\ensuremath{(6.113 \pm 1.829) \cdot 10^{-5}}}{6.113\cdot 10^{-5}}{1.829\cdot 10^{-5}}%
\htquantdef{Gamma944}{\Gamma_{944}}{\BRF{\tau^-}{2\pi^- \pi^+ \eta \nu_\tau (\eta \to \gamma\gamma) ~(\text{ex.~}K^0)}}{\ensuremath{(8.764 \pm 0.469) \cdot 10^{-5}}}{8.764\cdot 10^{-5}}{0.469\cdot 10^{-5}}%
\htquantdef{Gamma945}{\Gamma_{945}}{\BRF{\tau^-}{\pi^- 2\pi^0 \eta \nu_\tau}}{\ensuremath{(1.797 \pm 0.301) \cdot 10^{-4}}}{1.797\cdot 10^{-4}}{0.301\cdot 10^{-4}}%
\htquantdef{Gamma94by69}{\frac{\Gamma_{94}}{\Gamma_{69}}}{\frac{\BRF{\tau^-}{\pi^- K^- K^+ \pi^0 \nu_\tau}}{\BRF{\tau^-}{\pi^- \pi^- \pi^+ \pi^0 \nu_\tau\;(\text{ex.} K^0)}}}{\ensuremath{(0.1353 \pm 0.0406) \cdot 10^{-2}}}{0.1353\cdot 10^{-2}}{0.0406\cdot 10^{-2}}%
\htquantdef{Gamma96}{\Gamma_{96}}{\BRF{\tau^-}{K^- K^- K^+ \nu_\tau}}{\ensuremath{(2.156 \pm 0.800) \cdot 10^{-5}}}{2.156\cdot 10^{-5}}{0.800\cdot 10^{-5}}%
\htquantdef{Gamma998}{\Gamma_{998}}{1 - \Gamma_{\text{All}}}{\ensuremath{(9.858 \pm 9.849) \cdot 10^{-4}}}{9.858\cdot 10^{-4}}{9.849\cdot 10^{-4}}%
\htquantdef{Gamma9by5}{\frac{\Gamma_{9}}{\Gamma_{5}}}{\frac{\BRF{\tau^-}{\pi^- \nu_\tau}}{\BRF{\tau^-}{e^- \bar{\nu}_e \nu_\tau}}}{\ensuremath{0.6069 \pm 0.0032}}{0.6069}{0.0032}%
\htquantdef{GammaAll}{\Gamma_{\text{All}}}{\Gamma_{\text{All}}}{\ensuremath{0.99901 \pm 0.00098}}{0.99901}{0.00098}%
\htquantdef{gmubyge_tau}{gmubyge_tau}{}{\ensuremath{1.0018 \pm 0.0014}}{1.0018}{0.0014}%
\htquantdef{gtaubyge_tau}{gtaubyge_tau}{}{\ensuremath{1.0029 \pm 0.0015}}{1.0029}{0.0015}%
\htquantdef{gtaubygmu_fit}{gtaubygmu_fit}{}{\ensuremath{1.0001 \pm 0.0014}}{1.0001}{0.0014}%
\htquantdef{gtaubygmu_K}{gtaubygmu_K}{}{\ensuremath{0.9858 \pm 0.0071}}{0.9858}{0.0071}%
\htquantdef{gtaubygmu_pi}{gtaubygmu_pi}{}{\ensuremath{0.9963 \pm 0.0027}}{0.9963}{0.0027}%
\htquantdef{gtaubygmu_tau}{gtaubygmu_tau}{}{\ensuremath{1.0011 \pm 0.0015}}{1.0011}{0.0015}%
\htquantdef{hcut}{hcut}{}{\ensuremath{(6.58211928 \pm 0.00000015) \cdot 10^{-22}}}{6.58211928\cdot 10^{-22}}{0.00000015\cdot 10^{-22}}%
\htquantdef{KmKzsNu}{KmKzsNu}{}{7.950\cdot 10^{-4}}{7.950\cdot 10^{-4}}{0}%
\htquantdef{KmPizKzsNu}{KmPizKzsNu}{}{7.950\cdot 10^{-4}}{7.950\cdot 10^{-4}}{0}%
\htquantdef{KtoENu}{KtoENu}{}{\ensuremath{(1.5810 \pm 0.0080) \cdot 10^{-5}}}{1.5810\cdot 10^{-5}}{0.0080\cdot 10^{-5}}%
\htquantdef{KtoMuNu}{KtoMuNu}{}{\ensuremath{0.6355 \pm 0.0011}}{0.6355}{0.0011}%
\htquantdef{m_e}{m_e}{}{\ensuremath{0.510998928 \pm 0.000000011}}{0.510998928}{0.000000011}%
\htquantdef{m_K}{m_K}{}{\ensuremath{493.677 \pm 0.016}}{493.677}{0.016}%
\htquantdef{m_mu}{m_mu}{}{\ensuremath{105.6583715 \pm 0.0000035}}{105.6583715}{0.0000035}%
\htquantdef{m_pi}{m_pi}{}{\ensuremath{139.57018 \pm 0.00035}}{139.57018}{0.00035}%
\htquantdef{m_s}{m_s}{}{\ensuremath{93.50 \pm 2.50}}{93.50}{2.50}%
\htquantdef{m_tau}{m_tau}{}{\ensuremath{(1.77682 \pm 0.00016) \cdot 10^{3}}}{1.77682\cdot 10^{3}}{0.00016\cdot 10^{3}}%
\htquantdef{m_W}{m_W}{}{\ensuremath{(8.0399 \pm 0.0015) \cdot 10^{4}}}{8.0399\cdot 10^{4}}{0.0015\cdot 10^{4}}%
\htquantdef{phspf_mebymmu}{phspf_mebymmu}{}{\ensuremath{0.999812949174 \pm 0.000000000015}}{0.999812949174}{0.000000000015}%
\htquantdef{phspf_mebymtau}{phspf_mebymtau}{}{\ensuremath{0.99999933833 \pm 0.00000000012}}{0.99999933833}{0.00000000012}%
\htquantdef{phspf_mmubymtau}{phspf_mmubymtau}{}{\ensuremath{0.9725588 \pm 0.0000048}}{0.9725588}{0.0000048}%
\htquantdef{pi}{pi}{}{3.142}{3.142}{0}%
\htquantdef{PimKmKpNu}{PimKmKpNu}{}{0.1440\cdot 10^{-2}}{0.1440\cdot 10^{-2}}{0}%
\htquantdef{PimKmPipNu}{PimKmPipNu}{}{0.2940\cdot 10^{-2}}{0.2940\cdot 10^{-2}}{0}%
\htquantdef{PimKzsKzlNu}{PimKzsKzlNu}{}{9.774\cdot 10^{-4}}{9.774\cdot 10^{-4}}{0}%
\htquantdef{PimKzsKzsNu}{PimKzsKzsNu}{}{2.310\cdot 10^{-4}}{2.310\cdot 10^{-4}}{0}%
\htquantdef{PimPimPipNu}{PimPimPipNu}{}{9.020\cdot 10^{-2}}{9.020\cdot 10^{-2}}{0}%
\htquantdef{PimPimPipPizNu}{PimPimPipPizNu}{}{4.480\cdot 10^{-2}}{4.480\cdot 10^{-2}}{0}%
\htquantdef{PimPizKzsNu}{PimPizKzsNu}{}{0.1725\cdot 10^{-2}}{0.1725\cdot 10^{-2}}{0}%
\htquantdef{PimPizNu}{PimPizNu}{}{0.2552}{0.2552}{0}%
\htquantdef{pitoENu}{pitoENu}{}{\ensuremath{(1.2300 \pm 0.0040) \cdot 10^{-4}}}{1.2300\cdot 10^{-4}}{0.0040\cdot 10^{-4}}%
\htquantdef{pitoMuNu}{pitoMuNu}{}{\ensuremath{0.99987700 \pm 0.00000040}}{0.99987700}{0.00000040}%
\htquantdef{R_tau}{R_tau}{}{\ensuremath{3.6314 \pm 0.0081}}{3.6314}{0.0081}%
\htquantdef{R_tau_s}{R_tau_s}{}{\ensuremath{0.1618 \pm 0.0026}}{0.1618}{0.0026}%
\htquantdef{R_tau_VA}{R_tau_VA}{}{\ensuremath{3.4696 \pm 0.0080}}{3.4696}{0.0080}%
\htquantdef{rrad_LD_kmu_pimu}{rrad_LD_kmu_pimu}{}{\ensuremath{0.9930 \pm 0.0035}}{0.9930}{0.0035}%
\htquantdef{rrad_LD_tauK_taupi}{rrad_LD_tauK_taupi}{}{\ensuremath{1.0003 \pm 0.0044}}{1.0003}{0.0044}%
\htquantdef{rrad_tau_Knu}{rrad_tau_Knu}{}{\ensuremath{1.02010 \pm 0.00030}}{1.02010}{0.00030}%
\htquantdef{sigmataupmy4s}{sigmataupmy4s}{}{0.9190}{0.9190}{0}%
\htquantdef{tau_K}{tau_K}{}{\ensuremath{(1.2380 \pm 0.0021) \cdot 10^{-8}}}{1.2380\cdot 10^{-8}}{0.0021\cdot 10^{-8}}%
\htquantdef{tau_mu}{tau_mu}{}{\ensuremath{(2.196981 \pm 0.000022) \cdot 10^{-6}}}{2.196981\cdot 10^{-6}}{0.000022\cdot 10^{-6}}%
\htquantdef{tau_pi}{tau_pi}{}{\ensuremath{(2.60330 \pm 0.00050) \cdot 10^{-8}}}{2.60330\cdot 10^{-8}}{0.00050\cdot 10^{-8}}%
\htquantdef{tau_tau}{tau_tau}{}{290.3 \pm 0.5}{290.3}{0.5}%
\htquantdef{Vud}{Vud}{}{0.97425 \pm 0.00022}{0.97425}{0.00022}%
\htquantdef{Vud_moulson_ckm14}{Vud_moulson_ckm14}{}{\ensuremath{0.97417 \pm 0.00021}}{0.97417}{0.00021}%
\htquantdef{Vus}{Vus}{}{\ensuremath{0.2176 \pm 0.0021}}{0.2176}{0.0021}%
\htquantdef{Vus_err_exp}{Vus_err_exp}{}{0.1833\cdot 10^{-2}}{0.1833\cdot 10^{-2}}{0}%
\htquantdef{Vus_err_exp_perc}{Vus_err_exp_perc}{}{0.8425}{0.8425}{0}%
\htquantdef{Vus_err_perc}{Vus_err_perc}{}{0.9527}{0.9527}{0}%
\htquantdef{Vus_err_th}{Vus_err_th}{}{9.680\cdot 10^{-4}}{9.680\cdot 10^{-4}}{0}%
\htquantdef{Vus_err_th_perc}{Vus_err_th_perc}{}{0.44}{0.44}{0}%
\htquantdef{Vus_kl2_moulson_ckm14}{Vus_kl2_moulson_ckm14}{}{\ensuremath{0.22484 \pm 0.00059}}{0.22484}{0.00059}%
\htquantdef{Vus_kl3_moulson_ckm14}{Vus_kl3_moulson_ckm14}{}{\ensuremath{0.22320 \pm 0.00090}}{0.22320}{0.00090}%
\htquantdef{Vus_mism}{Vus_mism}{}{\ensuremath{(-0.7877 \pm 0.2304) \cdot 10^{-2}}}{-0.7877\cdot 10^{-2}}{0.2304\cdot 10^{-2}}%
\htquantdef{Vus_mism_sigma}{Vus_mism_sigma}{}{-3.4}{-3.4}{0}%
\htquantdef{Vus_mism_sigma_abs}{Vus_mism_sigma_abs}{}{3.4}{3.4}{0}%
\htquantdef{Vus_tau}{Vus_tau}{}{\ensuremath{0.2204 \pm 0.0014}}{0.2204}{0.0014}%
\htquantdef{Vus_tau_mism}{Vus_tau_mism}{}{\ensuremath{(-0.5105 \pm 0.1747) \cdot 10^{-2}}}{-0.5105\cdot 10^{-2}}{0.1747\cdot 10^{-2}}%
\htquantdef{Vus_tau_mism_sigma}{Vus_tau_mism_sigma}{}{-2.9}{-2.9}{0}%
\htquantdef{Vus_tau_mism_sigma_abs}{Vus_tau_mism_sigma_abs}{}{2.9}{2.9}{0}%
\htquantdef{Vus_tauKnu}{Vus_tauKnu}{}{\ensuremath{0.2212 \pm 0.0020}}{0.2212}{0.0020}%
\htquantdef{Vus_tauKnu_err_th_perc}{Vus_tauKnu_err_th_perc}{}{0.5760}{0.5760}{0}%
\htquantdef{Vus_tauKnu_mism}{Vus_tauKnu_mism}{}{\ensuremath{(-0.4238 \pm 0.2218) \cdot 10^{-2}}}{-0.4238\cdot 10^{-2}}{0.2218\cdot 10^{-2}}%
\htquantdef{Vus_tauKnu_mism_sigma}{Vus_tauKnu_mism_sigma}{}{-1.9}{-1.9}{0}%
\htquantdef{Vus_tauKnu_mism_sigma_abs}{Vus_tauKnu_mism_sigma_abs}{}{1.9}{1.9}{0}%
\htquantdef{Vus_tauKpi}{Vus_tauKpi}{}{\ensuremath{0.2232 \pm 0.0019}}{0.2232}{0.0019}%
\htquantdef{Vus_tauKpi_err_th_perc}{Vus_tauKpi_err_th_perc}{}{0.4731}{0.4731}{0}%
\htquantdef{Vus_tauKpi_err_th_perc_delta_LD_tauK_Kmu}{Vus_tauKpi_err_th_perc_delta_LD_tauK_Kmu}{}{-0.1090}{-0.1090}{0}%
\htquantdef{Vus_tauKpi_err_th_perc_delta_LD_taupi_pimu}{Vus_tauKpi_err_th_perc_delta_LD_taupi_pimu}{}{6.989\cdot 10^{-2}}{6.989\cdot 10^{-2}}{0}%
\htquantdef{Vus_tauKpi_err_th_perc_f_K_by_f_pi}{Vus_tauKpi_err_th_perc_f_K_by_f_pi}{}{-0.4195}{-0.4195}{0}%
\htquantdef{Vus_tauKpi_err_th_perc_rrad_LD_kmu_pimu}{Vus_tauKpi_err_th_perc_rrad_LD_kmu_pimu}{}{-0.1762}{-0.1762}{0}%
\htquantdef{Vus_tauKpi_mism}{Vus_tauKpi_mism}{}{\ensuremath{(-0.2281 \pm 0.2190) \cdot 10^{-2}}}{-0.2281\cdot 10^{-2}}{0.2190\cdot 10^{-2}}%
\htquantdef{Vus_tauKpi_mism_sigma}{Vus_tauKpi_mism_sigma}{}{-1.0}{-1.0}{0}%
\htquantdef{Vus_tauKpi_mism_sigma_abs}{Vus_tauKpi_mism_sigma_abs}{}{1.0}{1.0}{0}%
\htquantdef{Vus_uni}{Vus_uni}{}{\ensuremath{0.22547 \pm 0.00095}}{0.22547}{0.00095}%
\htquantdef{VusbyVud_moulson_ckm14}{VusbyVud_moulson_ckm14}{}{\ensuremath{0.23080 \pm 0.00060}}{0.23080}{0.00060}%
\htdef{couplingsCorr}{%
$\left( \frac{g_\tau}{g_e} \right)$ &   53\\
$\left( \frac{g_\mu}{g_e} \right)$ &  -49 &   48\\
$\left( \frac{g_\tau}{g_\mu} \right)_\pi$ &   24 &   26 &    2\\
$\left( \frac{g_\tau}{g_\mu} \right)_K$ &   12 &   10 &   -2 &    5\\
 & $\left( \frac{g_\tau}{g_\mu} \right)$ & $\left( \frac{g_\tau}{g_e} \right)$ & $\left( \frac{g_\mu}{g_e} \right)$ & $\left( \frac{g_\tau}{g_\mu} \right)_\pi$}%

\newenvironment{envsmall}%
  {\small}%
  {}
\newenvironment{ensuredisplaymath}
  {\(\displaystyle}
  {\)}

\newcommand{\BRF}[2]{#2}
\renewcommand{\BR}{\text{BR}\xspace}

\begin{document}
\begin{titlepage}
\pubblock

\vfill
\Title{Determination of \Vus from the \mtau lepton branching fractions}
\vfill
\Author{\paperAuthor}
\Address{\paperInst}
\vfill
\begin{Abstract}
We determine the Cabibbo-Kobayashi-Maskawa matrix (CKM) element \Vus in
several different ways using updated preliminary HFAG averages for the
\mtau lepton branching fractions and we compare the results with the
determinations obtained from the kaon decays and from the unitarity of the
CKM matrix.
\end{Abstract}
\vfill
\begin{Presented}
8th International Workshop on the CKM Unitarity Triangle (CKM 2014), Vienna, Austria, September 8--12, 2014
\end{Presented}
\vfill
\end{titlepage}
\def\thefootnote{\fnsymbol{footnote}}
\setcounter{footnote}{0}

\section{Introduction}

The CKM matrix element \Vus is most precisely determined from kaon
decays~\cite{Antonelli:2010yf}, and its precision is limited by the
uncertainties of the lattice QCD estimates of $f_+(0)$ and $f_K/f_\pi$.
Using the \mtau branching fractions it is possible to determine \Vus in an
alternative way~\cite{Gamiz:2006xx} that does not depend on lattice QCD and
has small theory uncertainties:
\begin{align}
  \Vus &= \sqrt{\Rstrange/\left[\frac{\Rnonstrange}{\Vud^2} - \delta R_{\text{theory}}\right]}~,\label{eq:vus-tau-incl}
\end{align}
where \Rstrange and \Rnonstrange are the ratio of the inclusive \mtau width
into strange and non-strange final states to the partial width into
electron ($\Gamma(\mtau \to e\nu_\tau\bar{\nu_e})$).
In this formula, the $SU(3)$ breaking parameter $\delta R_{\text{theory}}$
depends on the $s$ quark mass and is determined in the context of low
energy QCD theory, partly relying on experimental data on low energy QCD
processes~\cite{Gamiz:2006xx}. The required experimental data, including
uncertainties and correlations, are adequately provided by the HFAG
averages~\cite{Amhis:2012bh}. In the following, we update the HFAG averages
to September 2014 and we determine \Vus accordingly.
Furthermore, we use the updated HFAG data to determine \Vus with \mtau
decays in two additional ways, using $\BR(\tau\to K\nu)$. This provides an
alternative determination of \Vus that is independent of the kaon data but
is affected in a similar way by the parameters $f_K$ and $f_\pi$
(determined with lattice QCD). 

\section{\Vus from inclusive $\tau\to X_s\nu$}

Referring to Eq.~\ref{eq:vus-tau-incl}, we use preliminary HFAG averages
updated to September 2014 to compute the experimental inputs.
Following Ref.~\cite{Davier:2005xq}, we assume lepton universality to
obtain a the ``universality-improved'' $\BR_e = \BR(\tau \to
e\bar{\nu}_e\nu_\tau)$ using the \mtau branching
fraction to electron and muon and the \mtau lifetime, including the precise
2013 Belle result~\cite{Belous:2013dba}, obtaining $\BR_e^{\text{uni}} =
(\htuse{Be_univ})\%$. We get $\Rstrange = \BR(\tau\to X_s\nu) /
\BR_e^{\text{uni}} = \htuse{R_tau_s}$ (see also Table~\ref{tab:tau:vus}) and
$\Rnonstrange = \BR(\tau\to X_{\text{non-strange}}\nu) / \BR_e^{\text{uni}}
= \htuse{R_tau_VA}$. We update $\delta R_{\text{theory}}$ from
Ref.~\cite{Gamiz:2006xx} with the up-to-date value of the $s$-quark mass
$m_s = \htuse{m_s}$~\cite{PDG_2012} and get $\delta R_{\text{theory}} =
\htuse{deltaR_su3break}$. We chose not to use more recent calculations of
$\delta R_{\text{theory}}$~\cite{Gamiz:2007qs,Maltman:2010hb},
which have smaller or larger estimated
uncertainties. Using $\Vud = \htuse{Vud}$~\cite{PDG_2012},
we compute $\VusTauIncl = \htuse{Vus}$, which is
$\htuse{Vus_mism_sigma_abs}\sigma$ lower than the unitarity CKM prediction
$\VusUni = \htuse{Vus_uni}$, from $(\VusUni)^2 = 1 - \Vud^2$ (\Vub being
negligible). The
uncertainty contribution from $\delta R_{\text{theory}}$ is
\htuse{Vus_err_th_perc}\%.

\begin{table}[tb]
\begin{center}
\renewcommand*{\arraystretch}{1.3}%
\begin{envsmall}
\begin{center}
\begin{tabular}{llll}
\hline
\multicolumn{1}{c}{\bfseries Branching fraction} &
\multicolumn{1}{c}{\bfseries HFAG preliminary Summer 2014 fit} \\
\hline
\htuse{BrStrangeVal}
\\\hline
\htuse{BrStrangeTotVal}
\\\hline
\end{tabular}
\end{center}
\end{envsmall}
\caption{HFAG preliminary Summer 2014 \mtau branching fractions to strange final states.\label{tab:tau:vus}}%
\end{center}
\end{table}

\subsection{\Vus from $\BR(\tau \to K\pi\nu)$}

It is also possible to determine \Vus from the branching fraction $\BR(\tau
\to K\pi\nu)$~\cite{Antonelli:2013usa, Passemar:ckm:2014} from the equation:
\begin{align*}
  \Gamma(\tau \to \bar{K} \pi \nu_\tau [\gamma])
  =& \frac{G_F^2 m_\tau^5 }{96 \pi^3}\, C_K^{2} S_{\text{EW}}
  ^\tau\left(|V_{us}| \fpKpi\right)^2\times \\
  & \times I_{K}^\tau \left(1 + \dEM{K \tau} + \tdSU{K \pi} \right)^2~.
\end{align*}
The phase space integrals, $I_{K}^\ell$ are determined from the $K\pi$ form
factors. The first estimate of the long-distance electromagnetic
corrections ($\dEM{K \tau}$) have been computed~\cite{Antonelli:2013usa}. 
The isospin breaking corrections ($\tdSU{K \pi}$) allow using both the
measurement on $\tau \rightarrow K^-  \pi^0
\nu_\tau$ and $\tau \rightarrow K^0_S  \pi \nu_\tau$.
Using data from the HFAG 2012 fit, from kaon decays, \fpKpi from FLAG 2013~\cite{Aoki:2013ldr}, one obtains $\fpKpi\Vus =
0.2141 \pm 0.0021_{\text{exp}} \pm 0.0014_{I{K}\tau}$ and 
$\Vus = 0.2216 \pm 0.0027$~\cite{Passemar:ckm:2014}. Neglecting
correlations, this is $1.4\sigma$ below the CKM unitarity prediction.

\section{Conclusions}

\begin{figure}[tb]
\centering
\includegraphics[width=0.8\linewidth]{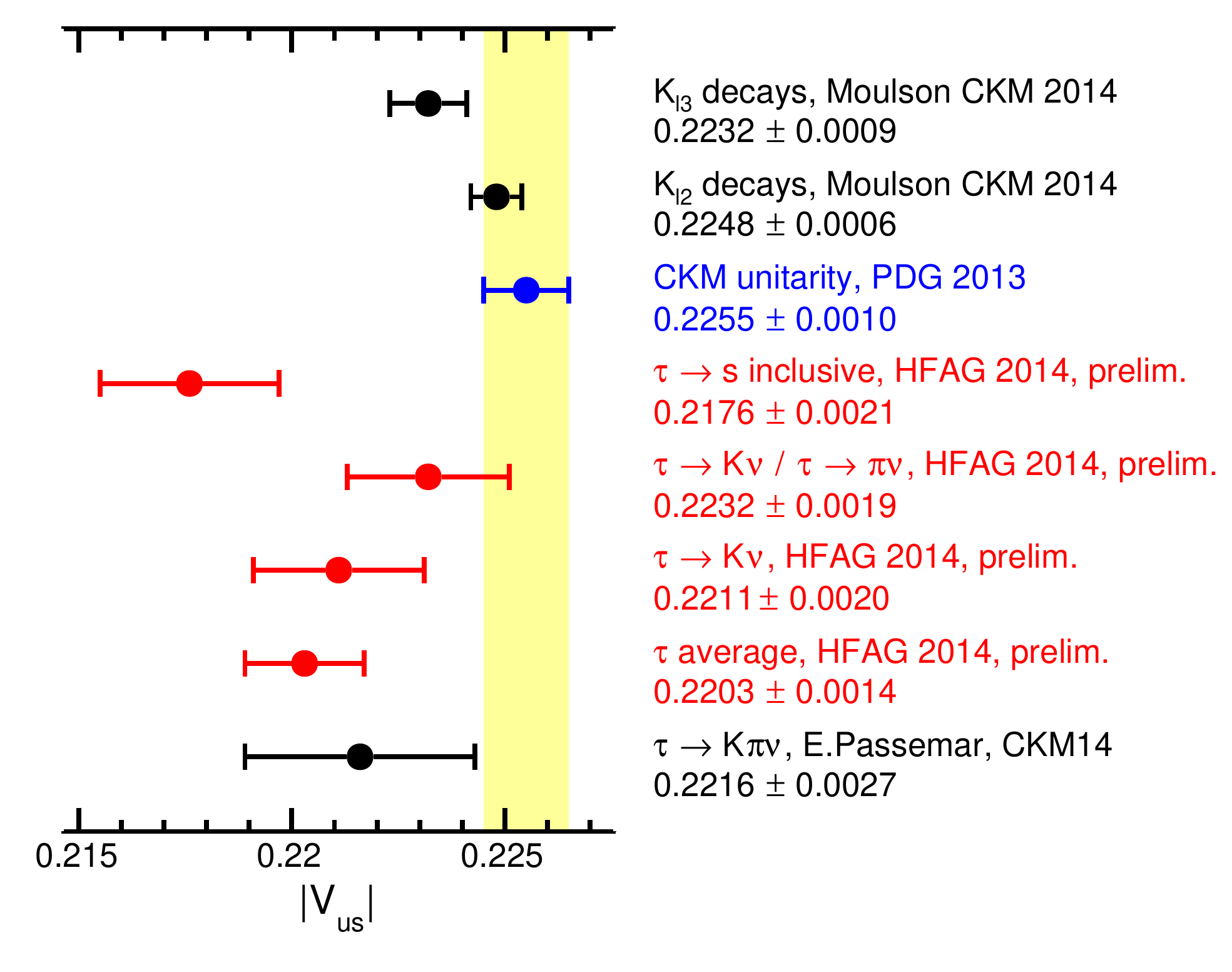}
\caption{\Vus from tau and kaon decays and from the CKM matrix unitarity.}
\label{fig:Vus}
\end{figure}

All determinations of \Vus from \mtau decays are lower than the CKM
matrix unitarity determination. In particular,
\Vus from inclusive $\tau\to X_s\nu$ is $\htuse{Vus_mism_sigma_abs}\sigma$
lower. Averaging the results using the HFAG 2014 preliminary data,
\VusTauIncl, \VusTauKpi and \VusTauKnu, we obtain:
\begin{alignat*}{6}
  & \Vus_\tau &&= \htuse{Vus_tau}~,
\end{alignat*}
$\htuse{Vus_tau_mism_sigma_abs}\sigma$ lower than the CKM unitarity prediction.
The correlation of the uncertainties on $f_K$ and $f_K/f_\pi$ is neglected,
since we could not find any estimate if it.
Even assuming $\pm 100\%$ correlation, the uncertainty on $\Vus_\tau$ does not change
more than about $\pm 5\%$.

The \Vus values obtained from kaon decays~\cite{Antonelli:2010yf}, on the
other hand, are statistically consistent with the unitarity determination.
Matthew Moulson has presented in the same conference updated and more
precise values of \Vus from kaon decays~\cite{Moulson:ckm:2014}, which
continue to be compatible with the unitarity
determination. Figure~\ref{fig:Vus} summarizes the \Vus results, using the
most up-to-date kaon results.


\ifdefined\bibtexflag
\bibliographystyle{h-physrev3-econf}%
\bibliography{%
  Summer14,%
  tau-refs-pdg,%
  tau-refs%
}%

\begin{thebibliography}{10}

\bibitem{Antonelli:2010yf}
M.~Antonelli {\em et~al.},
\newblock Eur. Phys. J. {\bf C69}, 399 (2010),
  \href{http://arxiv.org/abs/1005.2323}{{arXiv:1005.2323 [hep-ph]}}.

\bibitem{Gamiz:2006xx}
E.~Gamiz, M.~Jamin, A.~Pich, J.~Prades, and F.~Schwab,
\newblock Nucl. Phys. Proc. Suppl. {\bf 169}, 85 (2007),
  \href{http://arxiv.org/abs/hep-ph/0612154}{{arXiv:hep-ph/0612154}}.

\bibitem{Amhis:2012bh}
Y.~Amhis {\em et~al.},
\newblock (2012), \href{http://arxiv.org/abs/1207.1158}{{arXiv:1207.1158
  [hep-ex]}}.

\bibitem{Davier:2005xq}
M.~Davier, A.~Hocker, and Z.~Zhang,
\newblock Rev. Mod. Phys. {\bf 78}, 1043 (2006),
  \href{http://arxiv.org/abs/hep-ph/0507078}{{arXiv:hep-ph/0507078 [hep-ph]}}.

\bibitem{Belous:2013dba}
{Belle}, K.~Belous {\em et~al.},
\newblock Phys. Rev. Lett. {\bf 112}, 031801 (2014),
  \href{http://arxiv.org/abs/1310.8503}{{arXiv:1310.8503 [hep-ex]}}.

\bibitem{PDG_2012}
J.~Beringer {\em et~al.},
\newblock Phys. Rev. {\bf D86}, 010001 (2012),
\newblock and 2013 partial update for the 2014 edition.

\bibitem{Gamiz:2007qs}
E.~Gamiz, M.~Jamin, A.~Pich, J.~Prades, and F.~Schwab,
\newblock PoS {\bf KAON}, 008 (2008),
  \href{http://arxiv.org/abs/0709.0282}{{arXiv:0709.0282 [hep-ph]}}.

\bibitem{Maltman:2010hb}
K.~Maltman,
\newblock Nucl. Phys. Proc. Suppl. {\bf 218}, 146 (2011),
  \href{http://arxiv.org/abs/1011.6391}{{arXiv:1011.6391 [hep-ph]}}.

\bibitem{Antonelli:2013usa}
M.~Antonelli, V.~Cirigliano, A.~Lusiani, and E.~Passemar,
\newblock JHEP {\bf 1310}, 070 (2013),
  \href{http://arxiv.org/abs/1304.8134}{{arXiv:1304.8134 [hep-ph]}}.

\bibitem{Passemar:ckm:2014}
E.~Passemar,
\newblock {Determinations of $|V_{us}|$ from hadronic tau decays: A theory
  perspective},
\newblock {presented at the 8th International Workshop on the CKM Unitarity
  Triangle (CKM 2014), Vienna, Austria, September 8--12, 2014},
\newblock {to be published in the CKM 2014 conference proceedings}.

\bibitem{Aoki:2013ldr}
S.~Aoki {\em et~al.},
\newblock Eur. Phys. J. {\bf C74}, 2890 (2014),
  \href{http://arxiv.org/abs/1310.8555}{{arXiv:1310.8555 [hep-lat]}}.

\bibitem{Moulson:ckm:2014}
M.~Moulson,
\newblock {Experimental determination of $|V_{us}|$ from kaon decays},
\newblock {presented at the 8th International Workshop on the CKM Unitarity
  Triangle (CKM 2014), Vienna, Austria, September 8--12, 2014},
\newblock {to be published in the CKM 2014 conference proceedings}.

\end{thebibliography}
\else
\providecommand{\href}[2]{#2}%

\fi

\end{document}